\documentclass[10pt,aps,prx,twocolumn,superscriptaddress,nofootinbib]{revtex4-2}

\usepackage[utf8]{inputenc}
\usepackage[T1]{fontenc}
\usepackage{amsmath,amssymb,amsfonts,mathtools}
\usepackage{braket}
\usepackage{booktabs}
\usepackage{siunitx}
\usepackage{dcolumn}
\usepackage{graphicx}
\usepackage{float}
\usepackage[breaklinks=true,colorlinks,citecolor=blue,linkcolor=blue,urlcolor=blue]{hyperref}
\usepackage{microtype}
\usepackage{xcolor}
\usepackage{orcidlink}

\begin{document}


\title{Analog Quantum Feature Selection with Neutral-Atom Quantum Processors}

\author{José J. Orquín-Marqués$^{\orcidlink{0009-0004-5826-0910}}$}
\email{jose.j.orquin@uv.es}
\affiliation{IDAL, Electronic Engineering Department, ETSE-UV, University of Valencia, Avgda. Universitat s/n, 46100 Burjassot, Valencia, Spain}

\author{Carlos Flores-Garrig\'os$^{\orcidlink{0009-0000-9735-5411}}$}
\email{carlos.flores@kipu-quantum.com}
\affiliation{Kipu Quantum GmbH, Greifswalderstrasse 212, 10405 Berlin, Germany}
\affiliation{IDAL, Electronic Engineering Department, ETSE-UV, University of Valencia, Avgda. Universitat s/n, 46100 Burjassot, Valencia, Spain}

\author{Alejandro Gomez Cadavid$^{\orcidlink{0000-0001-8863-4806}}$}
\affiliation{Kipu Quantum GmbH, Greifswalderstrasse 212, 10405 Berlin, Germany}
\affiliation{Department of Physical Chemistry, University of the Basque Country UPV/EHU, Apartado 644, 48080 Bilbao, Spain}

\author{Anton Simen$^{\orcidlink{0000-0001-8863-4806}}$}
\affiliation{Kipu Quantum GmbH, Greifswalderstrasse 212, 10405 Berlin, Germany}
\affiliation{Department of Physical Chemistry, University of the Basque Country UPV/EHU, Apartado 644, 48080 Bilbao, Spain}

\author{Enrique Solano$^{\orcidlink{0000-0002-8602-1181}}$}
\affiliation{Kipu Quantum GmbH, Greifswalderstrasse 212, 10405 Berlin, Germany}

\author{Narendra N. Hegade$^{\orcidlink{0000-0002-9673-2833}}$}
\affiliation{Kipu Quantum GmbH, Greifswalderstrasse 212, 10405 Berlin, Germany}

\author{José D. Martín-Guerrero$^{\orcidlink{0000-0001-9378-0285}}$}
\email{jose.d.martin@uv.es}
\affiliation{IDAL, Electronic Engineering Department, ETSE-UV, University of Valencia, Avgda. Universitat s/n, 46100 Burjassot, Valencia, Spain}
\affiliation{Valencian Graduate School and Research Network of Artificial Intelligence (ValgrAI), Valencia, Spain}

\author{Yolanda Vives-Gilabert$^{\orcidlink{0000-0002-3744-5893}}$}
\affiliation{IDAL, Electronic Engineering Department, ETSE-UV, University of Valencia, Avgda. Universitat s/n, 46100 Burjassot, Valencia, Spain}

\date{\today}

\begin{abstract}
We present a quantum-native approach to quantum feature selection (QFS) based on analog quantum simulation with neutral atom arrays, adaptable to a variety of academic and industry applications. In our method, feature relevance—measured via mutual information with the target—is encoded as local detuning amplitudes, while feature redundancy is embedded through distance-dependent van der Waals interactions, constrained by the Rydberg blockade radius. The system is evolved adiabatically toward low-energy configurations, and the resulting measurement bitstrings are used to extract physically consistent subsets of features. The protocol is evaluated through simulation on three benchmark binary classification datasets: Adult Income, Bank Marketing, and Telco Churn. Compared to classical methods such as mutual information ranking and Boruta, combined with XGBoost and Random Forest classifiers, our quantum-computing approach achieves competitive or superior performance. In particular, for compact subsets, say 2–5 features, analog QFS improves mean AUC scores by 1.5–2.3\% while reducing the number of features by 75–84\%, offering interpretable, low-redundancy solutions. These results prove that programmable Rydberg arrays offer a viable platform for intelligent feature selection with practical relevance in machine learning pipelines, able to turn computational quantum advantage into industrial quantum usefulness.

\end{abstract}

\maketitle
\section{Introduction}

Feature selection (FS) is a crucial step in machine learning pipelines, aiming to identify informative and non-redundant variables that enhance predictive performance while reducing overfitting and computational cost~\cite{ISINGFORMULATION, FSPERSPECTIVE}. The task is inherently combinatorial and known to be NP-hard~\cite{QUBOFORM, SURVEYONFS}, leading classical approaches to rely on heuristics or greedy algorithms.

Quantum computing offers new possibilities for tackling such problems more efficiently, particularly through quantum annealing, variational algorithms, or analog Hamiltonian simulation~\cite{IndustryMIS, QSimCompRyd, ArbitraryConn}. In this work, we propose a quantum-computing FS method based on arrays of neutral atoms in Rydberg states, which support tunable long-range interactions and programmable spatial configurations~\cite{QCwithNA, AQC_NA, DAQLwithNA}.

We encode feature relevance via local detunings, based on mutual information with the target, and redundancy through pairwise Rydberg interactions modulated by interatomic distances. These distances are obtained from a multidimensional scaling (MDS) projection of the feature redundancy matrix~\cite{MDS, MDS_2}. The resulting system undergoes adiabatic evolution, ending in a ground state that reflects the optimal feature subset.

Our approach is implemented in simulation using Amazon Braket and evaluated on real-world classification datasets. Compared with classical methods such as Boruta and mutual information ranking~\cite{BORUTA, FS_REVIEW}, our quantum method shows competitive performance, particularly when selecting compact feature subsets—highlighting its potential in high-dimensional settings where classical search becomes intractable.

The rest of the paper is organized as follows:  
Section~\ref{sec:fs_ml} reviews the process of feature selection in classical machine learning.  
Section~\ref{sec:qfs_na} presents our quantum feature selection (QFS) method via neutral atom encoding.  
Section~\ref{sec:efficient_data_embedding} describes the data embedding process based on feature importance.  
Section~\ref{sec:evaluation} details the evaluation methodology, and Section~\ref{sec:results} reports the results.  
Finally, Section~\ref{sec:conclusions} provides the conclusions and perspectives for future work.

\section{Feature Selection in Machine Learning}
\label{sec:fs_ml}

Feature selection is a fundamental task in machine learning and data preprocessing, aiming to identify a subset of the most informative variables from a high-dimensional dataset. By retaining only the most relevant features, one can improve model interpretability, reduce overfitting, enhance generalization, and significantly decrease computational burden~\cite{INTROTOFS, FSPERSPECTIVE}.



Mutual information (MI) is one of the most popular metrics for feature selection due to its ability to capture nonlinear dependencies between a feature and the target variable. For a feature $X$ and class label $Y$, MI is usually represented as $I(X; Y)$. Features with higher mutual information values are more relevant for classification or regression tasks~\cite{SURVEYONFS}. However, selecting features based solely on individual relevance often leads to redundancy, where multiple features carry overlapping information. To address this, modern methods consider both relevance to the target and redundancy among features. The trade-off can be formalized as
\begin{equation}
\max_{z \in \{0,1\}^N}
\!\left[
\sum_i I(x_i;y)z_i
- \lambda \sum_{i<j} I(x_i;x_j)z_i z_j
\right],
\end{equation}
where $z_i$ indicates whether feature $i$ is selected and $\lambda$ balances relevance and redundancy.  
This formulation naturally maps to a \textit{Quadratic Unconstrained Binary Optimization} (QUBO) problem, solvable by classical heuristics or quantum-enhanced approaches~\cite{ISINGFORMULATION, QFS}.

Recent studies have explored quantum annealing, variational and evolutionary quantum algorithms, and analog quantum simulators to solve QUBO-based feature selection tasks more efficiently~\cite{QUBOFORM, SIMEN2023, QINSPIREDSWARM, QINSPIRED}. In this work, we exploit the analog simulation capabilities of neutral atom systems to implement a quantum-native feature selection scheme that encodes both mutual information and redundancy into the spatial arrangement and control parameters of a Rydberg atomic array.

\section{Neutral Atom Encoding for Quantum Feature Selection}
\label{sec:qfs_na}

Neutral atoms excited to Rydberg states have emerged as a promising hardware platform for quantum computation and simulation, particularly in analog quantum optimization. These systems combine excellent scalability, tunable long-range interactions, and flexible spatial programmability~\cite{Leveraging, EnhancedMeas, DAQLwithNA, FastSingleQ}. Individual atoms are confined by optical tweezers and coherently manipulated via laser pulses that drive transitions between the ground state $\ket{g}$ and an excited Rydberg state $\ket{r}$, forming an effective qubit.

Each atom behaves as a two-level system with $\ket{g}\!\equiv\!\ket{0}$ and $\ket{r}\!\equiv\!\ket{1}$.  
The time-dependent Hamiltonian for an array of $N$ atoms is
\begin{equation}
\begin{aligned}
H_{\text{Ryd}}(t) &=
\frac{\Omega(t)}{2}\sum_i
\!\left(
e^{i\phi(t)}\ket{g_i}\!\bra{r_i}
+\text{h.c.}
\right)
-\Delta_g(t)\sum_i \hat{n}_i \\
&\quad
-\Delta_l(t)\sum_i p_i \hat{n}_i
+\sum_{i<j} V_{ij}\hat{n}_i \hat{n}_j,
\end{aligned}
\label{RYD_H}
\end{equation}
where $\Omega(t)$ and $\phi(t)$ are the Rabi frequency and phase of the driving laser,  
$\Delta_g(t)$ and $\Delta_l(t)$ implement global and local detunings,  
$p_i$ encodes feature-dependent weights, and $V_{ij}$ are van der Waals interactions given by $V_{ij}=\frac{C_6}{|\vec{x}_i-\vec{x}_j|^6}$, with $C_6$ determined by the principal quantum number. The strong distance dependence of $V_{ij}$ leads to the \emph{Rydberg blockade}:  
if two atoms are closer than a critical distance $R_b$, their joint excitation is suppressed since $V_{ij}$ exceeds the laser bandwidth. The blockade radius is $R_b=\left(\frac{C_6}{\sqrt{\Omega^2+\Delta^2}}\right)^{1/6}$. This mechanism enforces spatial exclusion, making Rydberg arrays natural candidates for constraint-based optimization problems such as Maximum Independent Set (MIS)~\cite{EmbeddingMIS, OptProblemsRyd, StringBreaking}.  
Here, each atom represents a feature, and spatial arrangement encodes redundancy constraints.





Compared with superconducting qubits or trapped ions, neutral atoms offer scalability to hundreds of sites, flexible geometry, tunable interactions, and efficient fluorescence-based readout~\cite{Inter_Rydberg, Qinfo_Rydberg, GENERATIONPHASES, GeneticRyd}.

\subsection{Mapping Feature Selection to the Rydberg Hamiltonian}

The FS objective combining relevance and redundancy is defined as
\begin{equation}
Q(x;\alpha)=-\alpha \sum_i I_i x_i
+ (1-\alpha)\sum_{i<j} R_{ij} x_i x_j,
\end{equation}
where $I_i=I(x_i;y)$ and $R_{ij}=I(x_i;x_j)$.  
The final Hamiltonian becomes
\begin{equation}
H_{\text{Ryd}}(T)=
-\Delta_l^{\text{max}}\sum_i p_i\hat{n}_i
+\sum_{i<j} V_{ij}\hat{n}_i\hat{n}_j,
\end{equation}
linking feature relevance to detunings and redundancy to interaction strengths.

\subsection{Adiabatic Evolution Protocol}

\begin{figure}[H]
    \centering
    \includegraphics[width=0.95\linewidth]{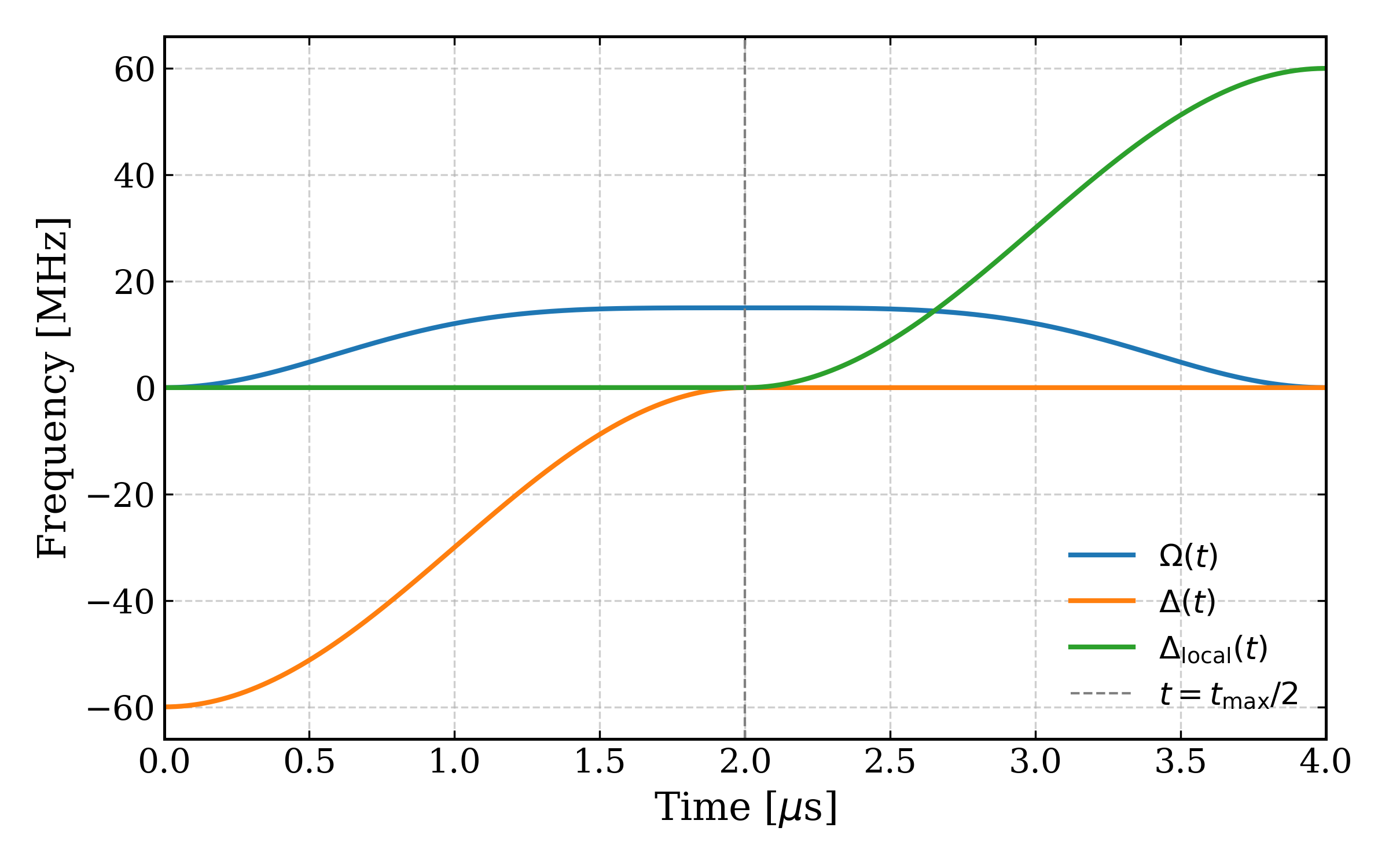}
    \caption{Driving profiles for selected schedules: the global detuning $\Delta_g(t)$ initializes the system, $\Omega(t)$ controls exploration, and $\Delta_l(t)$ encodes feature relevance.}
    \label{Drivings}
\end{figure}

Starting from all atoms in $\ket{g}$, we interpolate adiabatically toward the problem Hamiltonian.  
Schedules are designed to maintain adiabaticity while encoding relevance–redundancy balance. Fig.\ref{Drivings} shows the selected driving profiles for the driving protocol:

\begin{itemize}
\item $\Omega(t)$ ramps up and down smoothly for Hilbert-space exploration.
\item $\Delta_g(t)$ begins large and negative, then reduces to zero mid-protocol.
\item $\Delta_l(t)$ activates in the latter half, encoding $\{p_i\}$.
\end{itemize}

The slew rate, defined as
\begin{equation}
s(t)=\frac{1}{R_0}\frac{d\Delta(t)}{dt}, \quad R_0=\Omega^2/2\pi,
\end{equation}
quantifies the detuning sweep rate.  
Ensuring $|s(t)|\!\lesssim\!0.5$ suppresses diabatic transitions.  
These smooth drivings yield low-energy configurations approximating the optimal subset.

\section{Efficient Data Embedding Based on Feature Importance}
\label{sec:efficient_data_embedding}

To exploit the analog capabilities of neutral atom systems for feature selection, we propose a data embedding strategy that maps the mutual information structure of the dataset into the spatial configuration of an atomic array. This embedding reflects both feature relevance and pairwise redundancy in a physically meaningful way, consistent with the underlying Rydberg Hamiltonian.

\subsection{Mapping Feature Relationships to Physical Distances}

The first step encodes the mutual information between each feature and the target label, $I_i = I(x_i; y)$, into local detuning amplitudes applied to each atom. These determine the energetic bias for excitation and represent feature importance. Simultaneously, the pairwise mutual information $R_{ij} = I(x_i; x_j)$ between features defines spatial constraints: highly redundant features are placed closer together to strengthen Rydberg interactions, while weakly correlated features are kept apart.

We construct a theoretical distance matrix $D_{ij}$ derived from the redundancy matrix $R_{ij}$. For features $i$ and $j$, the raw (unnormalized) distance is
\begin{equation}
d_{ij} = \left(\frac{1}{R_{ij}}\right)^{1/6},
\end{equation}
ensuring that more redundant features are spatially closer. Distances are normalized to a physically meaningful interval,
\begin{equation}
d_{\min} = \frac{1}{\sqrt{2}} R_b, \quad d_{\max} = 4 R_b,
\end{equation}
where the blockade radius $R_b$ is
\begin{equation}
R_b = \left(\frac{C_6}{\Delta_l^{\text{max}2}}\right)^{1/6}.
\label{blockade_radius}
\end{equation}

At $d \approx d_{\min}$, excitations are energetically off-resonant, implementing a blockade constraint that penalizes redundant selections. At $d \approx d_{\max}$, interactions decay as $d^{-6}$ and features behave independently. Linearly normalizing $d_{ij}$ to $[d_{\min}, d_{\max}]$ yields the scaled matrix $D^{\text{scaled}}_{ij}$, ensuring a balance between physical feasibility and faithful encoding of feature correlations.

The redundancy structure is translated into spatial coordinates $\vec{x}_i \in \mathbb{R}^2$ through robust multidimensional scaling (MDS), producing an atomic layout that approximates the target distances $D^{\text{scaled}}_{ij}$ while enforcing the blockade constraint.

An adaptive interval $[d_{\min}, d_{\max}]$, initialized as $[R_b, 4R_b]$, guarantees that all atoms remain outside the blockade regime according to

\begin{equation}
\min_{i \neq j} \|\vec{x}_i - \vec{x}_j\| \ge R_b / \sqrt{2}.
\end{equation}

\begin{figure}[H]
    \centering
    \includegraphics[width=1.05\linewidth]{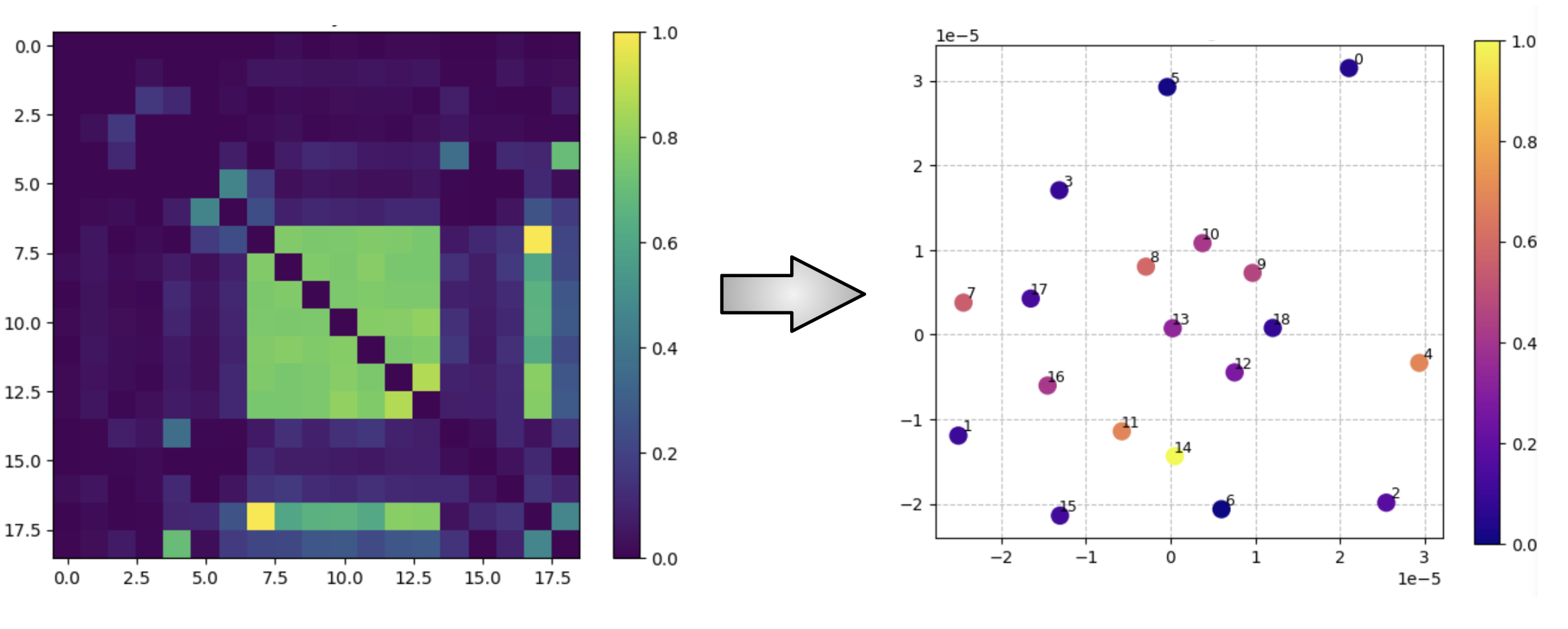}
    \caption{Mapping redundancy onto atom positions for one dataset. The algorithm places highly redundant features within blockade range while keeping independent features farther apart.}
    \label{Arrang}
\end{figure}

Because redundancy matrices are not strictly Euclidean, exact distance preservation is impossible. Therefore, multiple embeddings with different random seeds are computed, and the one with minimal reconstruction error between $D^{\text{scaled}}_{ij}$ and $\|\vec{x}_i - \vec{x}_j\|$ is selected. The relative error matrix
\begin{equation}
\epsilon_{ij} = \left| \frac{D^{\text{scaled}}_{ij} - \|\vec{x}_i - \vec{x}_j\|}{D^{\text{scaled}}_{ij}} \right|
\end{equation}
quantifies this distortion. In practice, average reconstruction errors $\varepsilon \lesssim 10^{-2}$–$10^{-1}$ are achieved, ensuring the embedding remains consistent with the physical constraints of the Rydberg platform. Fig. \ref{Arrang} shows the final arrangement for the Telco Churn dataset, after performing the iterative method of position embedding.

\subsection{Analog Quantum Simulation}

All simulations were carried out using the \texttt{Amazon Braket SDK} (\texttt{braket.ahs})~\cite{braket}, which provides a high-level programming interface for defining analog Hamiltonian protocols tailored to neutral atom platforms. The quantum evolution of the system was modeled using the Schrödinger equation solver backend, allowing for continuous-time simulation of the protocol in Eq.~(\ref{RYD_H}).

To implement the feature selection cost function, we made use of the \texttt{LocalDetuning} class to encode site-specific detuning amplitudes $\{p_i\}$ corresponding to feature relevance, while spatially embedded interactions modeled pairwise redundancies. The atomic positions and driving parameters define a time-dependent Hamiltonian that is smoothly interpolated from an initial easy-to-prepare state to a problem-specific final configuration. The evolution is discretized into 40 interpolation steps. The analog simulator integrates this evolution and samples the final state of the system across 10{,}000 repetitions (shots), enabling statistical analysis of measurement outcomes.

To capture realistic interaction effects, the Rydberg blockade radius $R_b$ is computed using Eq.\ref{blockade_radius} from the final detuning values. This parameter governs the effective interaction range between atoms, ensuring that the encoded redundancy constraints are faithfully enforced in the dynamics. Overall, this simulation framework provides a hardware-aligned and noise-free evaluation of the protocol’s performance, offering critical insights into the behavior of QFS prior to deployment on physical devices.

\subsection{Measurement, Post-Processing, and Solution Extraction}

At $t=T$, projective measurements in the computational basis collapse each atom into $\ket{g}$ or $\ket{r}$, corresponding to $x_i = 0$ or $x_i = 1$, respectively. Repeated measurements generate bitstrings representing candidate feature subsets. The lowest-energy $10\%$ of bitstrings under $Q(x;\alpha)$ are retained, focusing analysis on relevance–redundancy–balanced solutions.

To interpret the measurement outcomes, Rydberg densities $\langle \hat{n}_i \rangle$ are averaged across the low-energy ensemble. Since averaging can mask mutually exclusive combinations, a redundancy-aware post-processing is applied to produce multiple valid subsets and the best subset for each cardinality. Features are ranked by average excitation probability, and redundant features (correlation above a fixed threshold, e.g., 0.7) are pruned iteratively to generate complementary, non-overlapping sets. This filtering yields both a family of diverse redundancy-aware alternatives and a single optimal subset for any given feature count, ensuring interpretability and robustness in subsequent classification tasks.

\section{Performance Evaluation}
\label{sec:evaluation}

\subsection{Datasets}

The proposed analog QFS framework was evaluated on three publicly available binary classification datasets spanning telecommunications, socioeconomics, and marketing: \textbf{Adult Income} \cite{adult_2}, \textbf{Bank Marketing} \cite{bank_marketing_222}, and \textbf{Telco Churn} \cite{churn}. These datasets include a mix of categorical and numerical variables and exhibit diverse redundancy structures.

All categorical variables were transformed into integer codes via label encoding, and missing entries were removed or imputed as needed. Numerical features were standardized to zero mean and unit variance. Table~\ref{tab:datasets} summarizes the dimensionality and classes of each dataset. 

\begin{table}[H]
\centering
\caption{Summary of datasets used in the experiments.}
\label{tab:datasets}
\begin{tabular}{lccc}
\toprule
\textbf{Dataset} & \textbf{Samples} & \textbf{Features} & \textbf{Classes} \\
\midrule
Adult Income & 48,842 & 14 & 2 \\
Bank Marketing & 45,211 & 16 & 2 \\
Telco Customer Churn & 7,043 & 19 & 2 \\
\bottomrule
\end{tabular}
\end{table}

\subsection{Evaluation Metrics and Classification Procedure}

To assess predictive performance, we trained two supervised classifiers—\texttt{XGBoost}~\cite{ar:XGBoost} and \texttt{Random Forest}~\cite{RF}—using only the selected features. Both models underwent hyperparameter optimization via grid and randomized search with cross-validation. This ensures that the reported scores reflect model-independent relevance of the selected features.

Each dataset was split into training and test sets, and the following metrics were computed on the test partition:
Area Under the ROC Curve (AUC), precision, and recall.  
Five random seeds were used to account for stochasticity, and results were aggregated across the best-performing configurations.

\subsection{Classical Baselines}

Two classical baselines were used as a means for comparison:  
(i) \textbf{Mutual Information Ranking}, which ranks features by $I(x_i;y)$ and selects the top $k$; and  
(ii) \textbf{Boruta}~\cite{BORUTA}, a wrapper method that retains only statistically significant features compared to randomized shadow features.  
These baselines represent, respectively, relevance-driven and statistical robustness–driven selection strategies.

\newpage

\section{Results}
\label{sec:results}

We evaluated QFS against the classical baselines on all datasets. The atomic embedding successfully produced physically feasible configurations consistent with redundancy constraints: highly redundant features were mapped within blockade range, while weakly correlated ones were positioned farther apart. This demonstrates the ability of the multidimensional scaling procedure to project statistical relationships into a realizable spatial geometry.

\begin{figure}[H]
    \centering
    \includegraphics[width=0.5\textwidth]{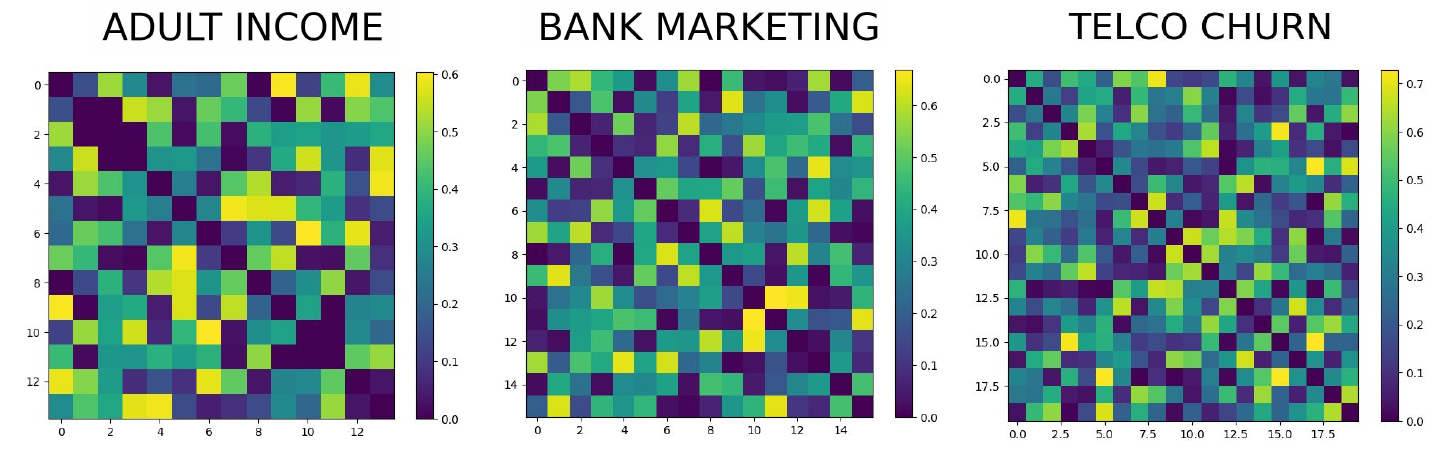}
    \caption{Relative error matrices $\epsilon_{ij}$ for the MDS embeddings of (left) Adult, (center) Bank Marketing, and (right) Telco Churn. Darker colors indicate smaller deviation between target and embedded distances.}
    \label{error_matrices}
\end{figure}

To quantify embedding quality, error matrix are represented in \ref{error_matrices}. We computed the mean relative error between the target (rescaled) distance matrix and the resulting interatomic distances.  Across datasets, mean errors were $\langle \epsilon_{ij} \rangle \approx 0.24$ for Adult, $\approx 0.26$ for Bank Marketing, and $\approx 0.25$ for Telco Churn, confirming that the geometric fidelity of the MDS mapping remains high despite the non-Euclidean structure of redundancy matrices. 

\begin{figure*}[t]
    \centering
    \includegraphics[width=0.95\textwidth]{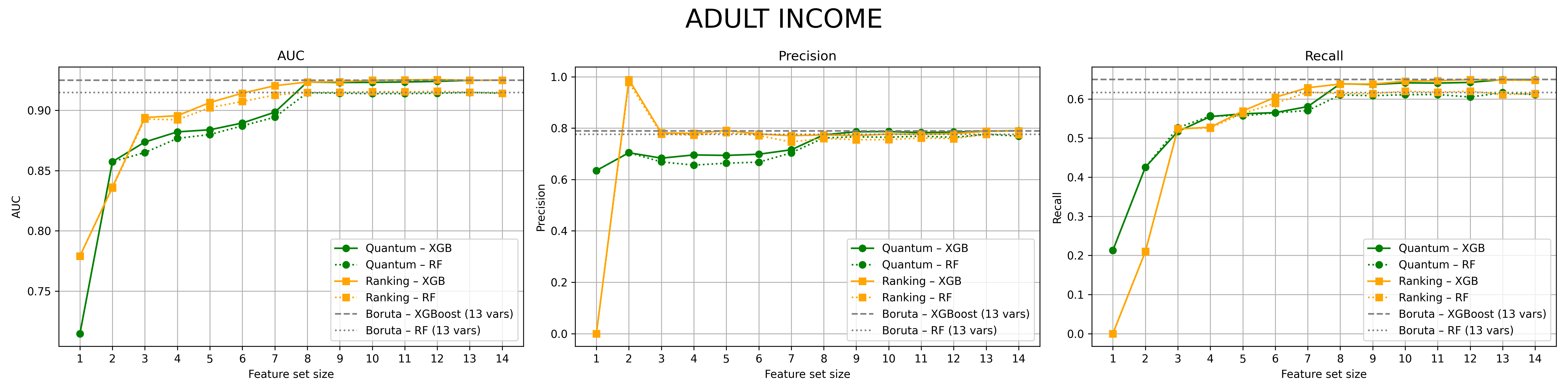}
    \includegraphics[width=0.95\textwidth]{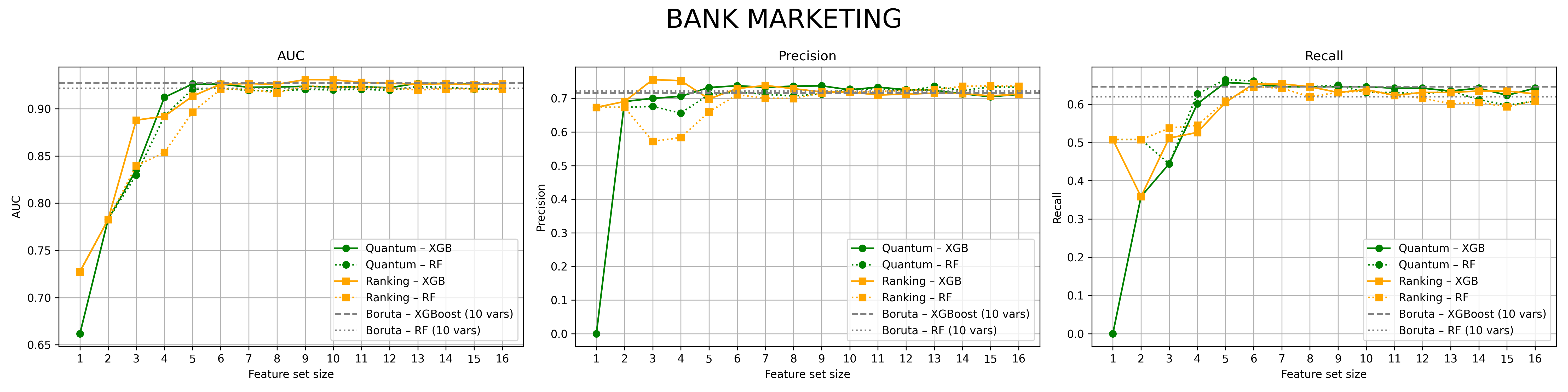}
    \includegraphics[width=0.95\textwidth]{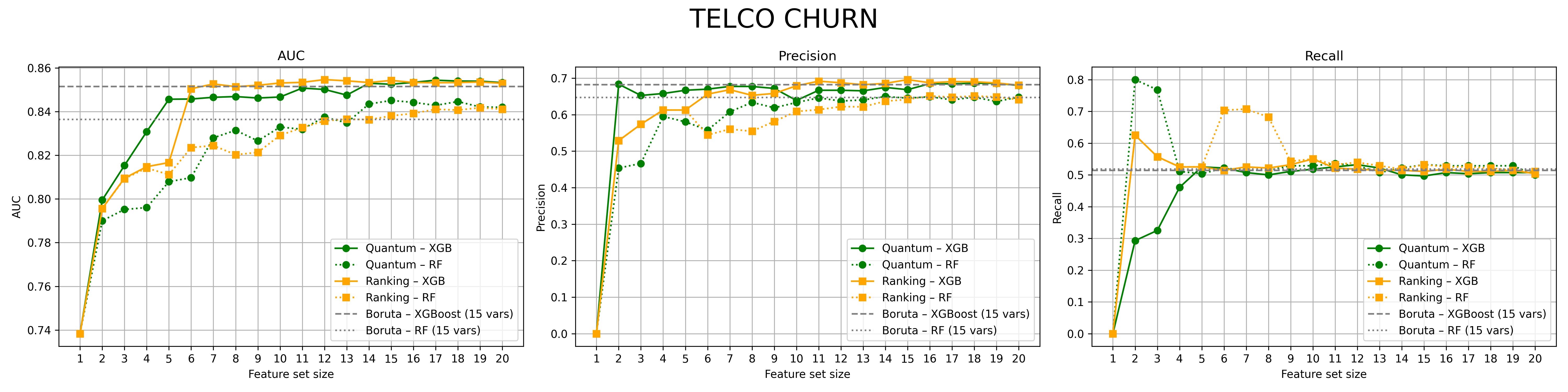}
    \caption{Median values of AUC, precision, and recall across the three datasets as a function of the number of selected features ($n_{\text{var}}$). The plots highlight that quantum feature selection (QFS) performs particularly well in the early stages of variable selection, achieving strong classification performance with small subsets of features.}
    \label{plot_metrics}
\end{figure*}

We compared the predictive performance of QFS and classical methods for feature subsets of sizes $1$–$6$. Resulting median values of the selected metrics are represented for all sizes of subsets in Fig.~\ref{plot_metrics}.
Table~\ref{table_metrics} reports median values of AUC, precision, and recall across all datasets for top-performing quantum subsets.

\begin{table}[H]
\caption{Comparison of median AUC, precision, and recall for selected feature subsets for each dataset, after performing the classification with XGBoost model. Bold AUC indicates superior QFS performance relative to the classical ranking baseline.}
\label{table_metrics}
\centering
\begin{tabular}{lccccc}
\toprule
\textbf{Dataset} & \textbf{Features} & \textbf{Method} & \textbf{AUC} & \textbf{Precision} & \textbf{Recall} \\
\midrule
Adult & 2 & QFS & \textbf{0.857} & 0.704 & \textbf{0.425} \\
      &   & Ranking & 0.836 & 0.989 & 0.209 \\
Bank  & 4 & QFS & \textbf{0.912} & 0.706 & \textbf{0.601} \\
      &   & Ranking & 0.892 & 0.753 & 0.526 \\
Churn & 3 & QFS & \textbf{0.815} & \textbf{0.652} & 0.325 \\
      &   & Ranking & 0.809 & 0.574 & \textbf{0.557} \\
\bottomrule
\end{tabular}
\end{table}

\begin{figure}[H]
    \centering
    \includegraphics[width=0.80\linewidth]{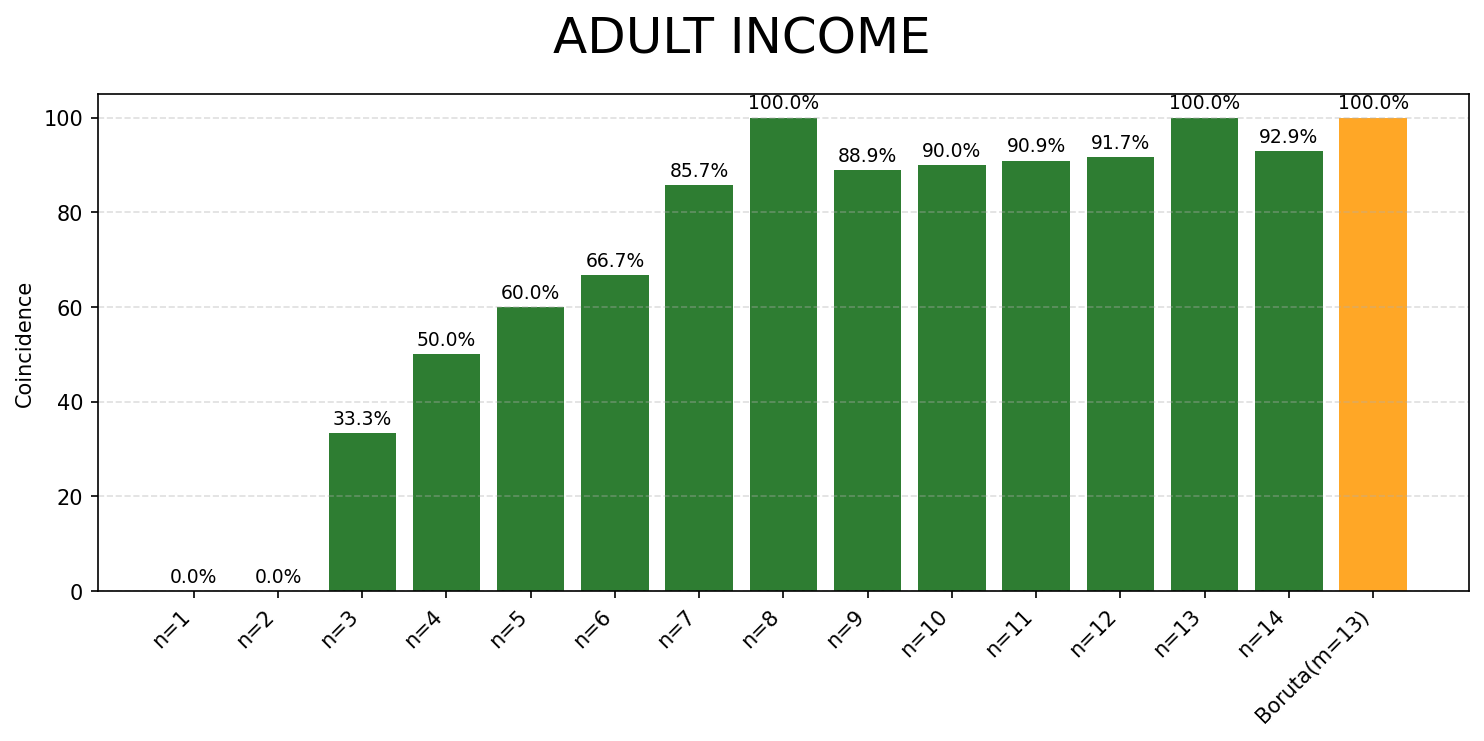}
    \vspace{0.2cm}
    \includegraphics[width=0.80\linewidth]{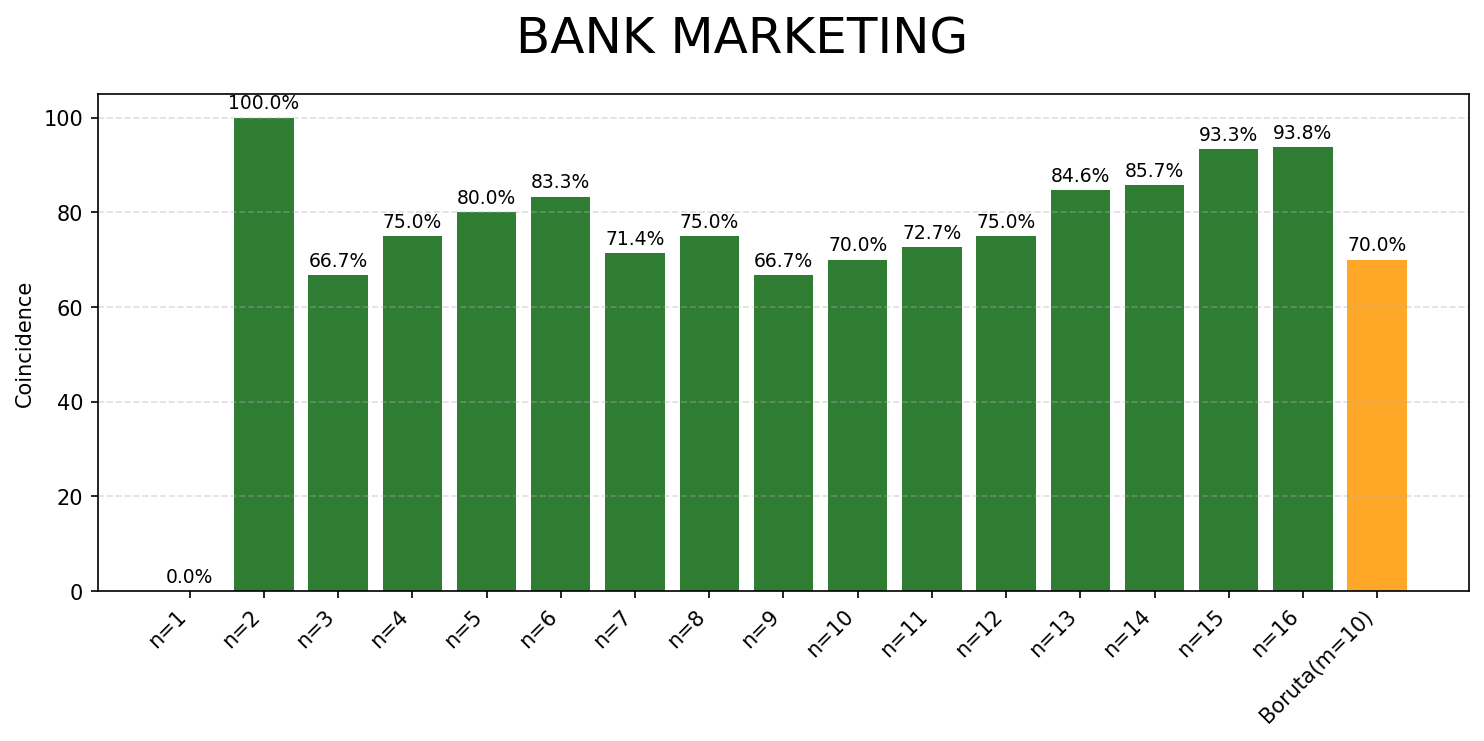}
    \vspace{0.2cm}
    \includegraphics[width=0.80\linewidth]{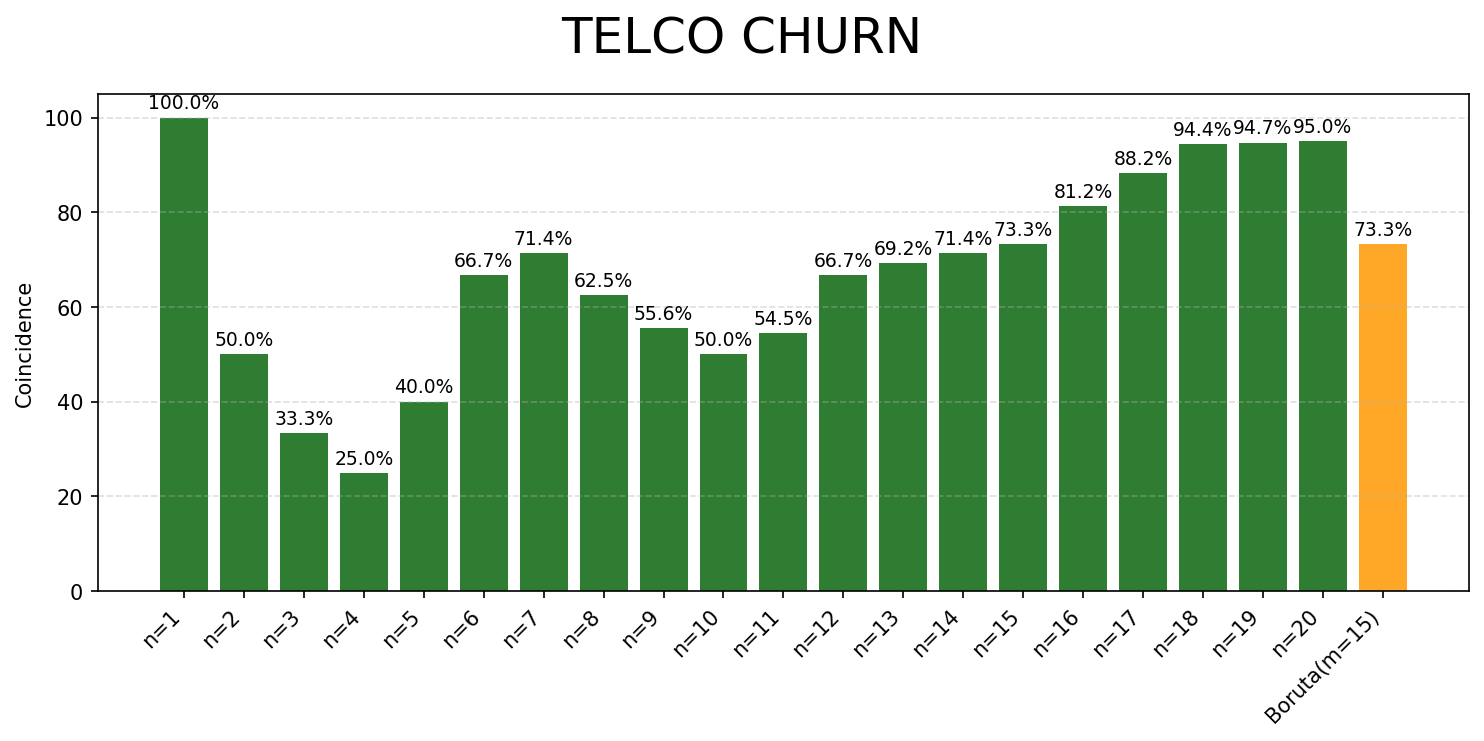}
    \caption{Percentage of coincident features between QFS-selected subsets and classical baselines as a function of subset size $n$.}
    \label{fig:overlap_plots}
\end{figure}

Figure~\ref{fig:overlap_plots} shows the overlap between QFS and classical subsets.  
For large subset sizes, overlap exceeds $80\%$–$90\%$, while smaller subsets show greater divergence—precisely where QFS often achieves superior AUC performance.  
This indicates that QFS can identify compact, non-obvious, and redundancy-balanced feature combinations not captured by standard relevance ranking.

\section{Conclusions and Outlook}
\label{sec:conclusions}

In this work, we introduced a quantum-native strategy for feature selection based on analog simulation with neutral atom arrays. By encoding mutual information and feature redundancy into a time-dependent Rydberg Hamiltonian, we framed feature selection as a physically grounded optimization problem governed by programmable interactions and detunings.

The proposed protocol uses adiabatic evolution to prepare low-energy configurations that balance individual relevance with pairwise redundancy. After evolution, we extract compact, interpretable feature subsets by analyzing the Rydberg excitation statistics and applying a redundancy-aware post-processing procedure. This approach requires minimal heuristic design and can yield multiple high-quality candidate subsets.

Our empirical results across three real-world binary classification datasets demonstrate that Quantum Feature Selection (QFS) achieves competitive or superior performance compared to strong classical baselines. In particular, QFS shows a consistent advantage when selecting small subsets (2–4 features), where classical methods are more susceptible to redundancy and instability. These compact selections are especially important in resource-constrained scenarios or in domains requiring interpretable, low-complexity models.

Importantly, although this study focuses on problems with up to 20 features, the QFS framework is inherently scalable. The analog quantum model supports a natural encoding of larger feature spaces by expanding the atomic array, and the blockade-induced constraints provide an effective mechanism to filter redundant information even in high-dimensional regimes. In such settings, where classical combinatorial search becomes intractable, QFS offers a physically efficient alternative capable of navigating the exponentially large solution space.

While the current implementation relies on simulation using Amazon Braket’s \texttt{ahs} environment, the method is compatible with near-term Rydberg hardware. As coherence times, spatial resolution, and control precision continue to improve, QFS can be extended to significantly larger systems, enabling practical quantum-assisted preprocessing in real-world data pipelines.

Future work may explore the inclusion of dynamic detuning optimization, extensions to multiclass or unsupervised learning settings, and integration with hybrid quantum–classical architectures. Overall, analog QFS emerges as a physically interpretable, statistically principled, and computationally scalable approach to feature selection, with strong potential for deployment in next-generation quantum machine learning workflows.

\section{Acknowledgment}

This work has been partially supported by the Valencian Government Grant No. CIAICO/2024/111 and the Spanish Ministry of Economic Affairs and Digital Transformation through the QUANTUM ENIA project call – Quantum Spain project, and the European Union through the Recovery, Trans- formation and Resilience Plan – NextGenerationEU (Digital Spain 2026) Agenda.
\newpage

\bibliography{bibliografia}

\end{document}